\documentclass[twocolumn]{aastex63}
\usepackage{multirow}
\usepackage{amsmath}
\usepackage{booktabs}

\usepackage{xcolor}

\newcommand\lsim{\mathrel{\rlap{\lower4pt\hbox{\hskip1pt$\sim$}}
\raise1pt\hbox{$<$}}}
\accepted{ApJL}
\shorttitle{Star Formation in SIGOs}
\shortauthors{Lake et al.}
\graphicspath{{./}{figures/}}

\begin{document}

\title{ The Supersonic Project: Star Formation in Early Star Clusters without Dark Matter}


\correspondingauthor{William Lake}
\email{wlake@astro.ucla.edu}
\author[0000-0002-4227-7919]{William Lake}
\affil{Department of Physics and Astronomy, UCLA, Los Angeles, CA 90095}
\affil{Mani L. Bhaumik Institute for Theoretical Physics, Department of Physics and Astronomy, UCLA, Los Angeles, CA 90095, USA\\}

\author[0000-0002-9802-9279]{Smadar Naoz}
\affil{Department of Physics and Astronomy, UCLA, Los Angeles, CA 90095}
\affil{Mani L. Bhaumik Institute for Theoretical Physics, Department of Physics and Astronomy, UCLA, Los Angeles, CA 90095, USA\\}

\author[0000-0003-3816-7028]{Federico Marinacci}
\affiliation{Department of Physics \& Astronomy ``Augusto Righi", University of Bologna, via Gobetti 93/2, 40129 Bologna, Italy\\}

\author[0000-0001-5817-5944]{Blakesley Burkhart}
\affiliation{Department of Physics and Astronomy, Rutgers, The State University of New Jersey, 136 Frelinghuysen Rd, Piscataway, NJ 08854, USA \\}
\affiliation{Center for Computational Astrophysics, Flatiron Institute, 162 Fifth Avenue, New York, NY 10010, USA \\}

\author[0000-0001-8593-7692]{Mark Vogelsberger}
\affil{Department of Physics and Kavli Institute for Astrophysics and Space Research, Massachusetts Institute of Technology, Cambridge, MA 02139, USA\\}

\author[0000-0003-2369-2911]{Claire E. Williams}
\affil{Department of Physics and Astronomy, UCLA, Los Angeles, CA 90095}
\affil{Mani L. Bhaumik Institute for Theoretical Physics, Department of Physics and Astronomy, UCLA, Los Angeles, CA 90095, USA\\}

\author[0000-0003-4962-5768]{Yeou S. Chiou}
\affil{Department of Physics and Astronomy, UCLA, Los Angeles, CA 90095}
\affil{Mani L. Bhaumik Institute for Theoretical Physics, Department of Physics and Astronomy, UCLA, Los Angeles, CA 90095, USA\\}

\author[0000-0001-6246-2866]{Gen Chiaki}
\affiliation{Astronomical Institute, Tohoku University, 6-3, Aramaki, Aoba-ku, Sendai, Miyagi 980-8578, Japan}

\author[0000-0002-0984-7713]{Yurina Nakazato}
\affiliation{Department of Physics, The University of Tokyo, 7-3-1 Hongo, Bunkyo, Tokyo 113-0033, Japan}

\author[0000-0001-7925-238X]{Naoki Yoshida}
\affiliation{Department of Physics, The University of Tokyo, 7-3-1 Hongo, Bunkyo, Tokyo 113-0033, Japan}
\affiliation{Kavli Institute for the Physics and Mathematics of the Universe (WPI), UT Institute for Advanced Study, The University of Tokyo, Kashiwa, Chiba 277-8583, Japan}
\affiliation{Research Center for the Early Universe, School of Science, The University of Tokyo, 7-3-1 Hongo, Bunkyo, Tokyo 113-0033, Japan}



\begin{abstract}

The formation mechanism of globular clusters (GCs) has long been debated by astronomers. It was recently proposed that Supersonically Induced Gas Objects (SIGOs)--which formed in the early Universe due to the supersonic relative motion of baryons and dark matter at recombination--could be the progenitors of early globular clusters. In order to become GCs, SIGOs must form stars relatively efficiently despite forming outside of dark matter halos. We investigate the potential for star formation in SIGOs using cosmological hydrodynamic simulations, including the aforementioned relative motions of baryons and dark matter,  molecular hydrogen cooling in primordial gas clouds, and explicit star formation. We find that SIGOs do form stars and that the nascent star clusters formed through this process are accreted by dark matter halos on short timescales ($\sim$ a few hundred Myr). Thus, SIGOs may be found as intact substructures within these halos, analogous to many present-day GCs. From this result, we conclude that SIGOs are capable of forming star clusters with similar properties to globular clusters in the early Universe and we discuss their detectablity by upcoming JWST surveys.

\end{abstract}

\keywords{Globular star clusters — High-redshift galaxies — Star formation
— Galactic and extragalactic astronomy}

\section{Introduction}

Globular clusters (GCs) are very old \citep[$\sim13$ Gyr,][]{Trenti+15} and dense structures whose formation mechanism has long been debated \citep[see e.g.,][]{Gunn80,Peebles84, Ashman+92,Harris+94,Kravtsov+05,Mashchenko+05I,Saitoh+06,Gray+11,Bekki+12,Kruijssen2015,Mandelker+18}. Observations indicate that GCs are likely enriched in baryons relative to the overall Universe \citep[e.g.,][]{Heggie+96,Bradford+11,Conroy+11,Ibata+13}, and that older GCs possess properties that may distinguish them from younger GCs \citep[see for a review][]{Bastian+18}. These properties create some uncertainty regarding the formation of GCs within the hierarchical picture of structure formation. To address this uncertainty, several formation scenarios have been proposed in the literature. 

One such formation mechanism for globular clusters posits that they formed at the high-efficiency end of normal galactic star formation, evolving from particularly dense giant molecular clouds (GMCs) \citep[e.g.,][]{Elmegreen+97,Kravtsov+05,Shapiro2010,Grudic+22}. This is supported by observations of massive young clusters in the merging Antennae system \citep[][]{Whitmore+95,Whitmore+99}. This picture is attractive in part because it naturally explains why GCs tend to be very old, from a time in the Universe when these particularly dense GMCs were more common. However, it is not obvious whether the age distribution of GCs from this model is compatible with observations, nor why the GC luminosity function appears similar across environments given this model.

A second theory suggests that GCs form inside DM halos, as suggested by \citet{Peebles84}, which were then stripped by the tidal field of their present-day host galaxies \citep[e.g.,][]{Bromm+02, Mashchenko+05I, Saitoh+06, Bekki+12,Donkelaar+23}. The primary strength of this theory is its natural connection between the properties (for example, the total mass) of a galaxy's GCs with its dark matter. It intuitively explains the scaling of these properties with the halo mass, and explains GC ages. However, this theory struggles to explain the observed presence of stellar tidal tails from some GCs, as the extended DM halos this model predicts should help to shield the formed clusters from tidal effects \citep[e.g.,][]{Grillmair+95,Moore96,Odenkirchen+03, Mashchenko+05I}. 

In this work, we explore star formation within a third formation mechanism for GCs proposed by \citet{naoznarayan14}. In this theory, at the time of recombination, as baryons decoupled from the photon field and cooled, the average sound speed in the Universe dropped precipitously. This drop caused the average relative velocity\footnote{Also known as the streaming velocity due to its coherence on few-Mpc scales} between baryons and DM in the Universe (about $30$ km s$^{-1}$) to become highly supersonic \citep[][]{TH,tseliakhovich11}. Following recombination, in the standard model of structure formation, baryon overdensities began to collapse, driven by existing DM overdensities that by this time were about $10^5$ times larger than baryon overdensities \citep[e.g.][]{NB}. 
The significant relative velocity between baryons and DM complicated this process. \citet{naoznarayan14} showed analytically that a sufficiently large relative velocity between baryons and their parent DM halo would create a spatial offset between the collapsing baryon overdensity and its parent halo. This effect is particularly important to understanding our local Universe, as \citet{Uysal+22} recently estimated that our Local Group formed in a region of the Universe with a high ($\sim 1.7\sigma$) value of the streaming velocity. In certain instances (especially at high gas masses, such as M$_{\rm gas} > {\rm few} \times 10^6$ M$_\odot$), the spatial offset produced by the effect is smaller than the virial radius of the parent DM halo, leading to an offset between the centers of gas and DM within halos. The resulting objects are known as Dark Matter + Gas Halos Offset by Streaming \citep[DM GHOSTs][]{Williams+22}, and have unique morphological and kinematic properties. 

In other instances, especially when M$_{\rm gas}\lesssim {\rm few} \times 10^6$ M$_\odot$, the spatial offset between DM and gas within these overdensities is large enough that the baryon overdensity collapses outside the virial radius of its parent halo. \citet{naoznarayan14} showed that this process would create a gas object with a characteristic mass of $10^4$ -- a few $\times$ $10^6$ M$_\odot$, which would be depleted of dark matter. This would put the formed objects squarely in the mass range of globular clusters, and given their early Universe nature and presumably low metallicities (any metals they have before star formation must originate from pollution from nearby halos), is suggestive of a connection to the low-metallicity sub-population of GCs \citep{Lake+21}. As SIGOs form from pristine gas, this also could lead to star clusters formed partially or entirely of Population III stars. Because our Local Group likely formed in a region with a large streaming velocity, the evolved forms of these objects are theoretically expected to be present in the Milky Way \citep{Uysal+22}.

Follow-up studies of these objects--known as Supersonically Induced Gas Objects (or SIGOs)--found them in simulations \citep[][]{popa,chiou18,chiou+19,Chiou+21,Lake+21}, and predicted distinctive observational signals from these objects \citep{Lake+21}. However, the connection between SIGOs and GCs is still not firmly established and depends in no small part on the star formation efficiency and stellar properties of SIGOs. Work by \citet{Nakazoto+22} using hydrodynamic simulations established that SIGOs are indeed capable of forming stars, following one such SIGO in a zoom-in simulation to Jeans collapse and demonstrating that molecular hydrogen cooling is sufficient to form stars in SIGOs. \citet{Lake+22} expanded upon this, providing initial estimates of the abundance of star-forming SIGOs and of the timescales important to their ability to form stars outside of halos. However, this work left open many questions about the properties of star-forming SIGOs, such as the efficiency of star formation in SIGOs that do form stars, and the fraction of SIGOs that form stars at redshifts later than ${\rm z}=20$.

When considering star formation in SIGOs, it is vital to consider molecular hydrogen cooling \citep[e.g.,][]{Glover13, Schauer+21, Nakazoto+22}. H$_2$ cooling allows the temperature of primordial gas clouds to lower to $\sim200$ K, which lowers their Jeans masses to $\sim1000$ M$_\odot$, potentially allowing these primordial gas clouds to collapse and form stars \citep{Yoshida+08}. With these factors in mind, in the present \textit{letter} we present the results of a suite of {\tt AREPO} simulations including the streaming velocity and incorporating molecular hydrogen cooling, with the aim of constraining some of the properties of star-forming SIGOs, including their star-formation efficiency.

This \textit{letter} is organized as follows: in Section \ref{sec:Methods} we detail the simulation setup. In Section \ref{sec:Results} we discuss the bulk properties of star formation in SIGOs, as well as comparing star formation in SIGOs to more classical star formation within DM halos. Lastly, in Section \ref{sec:Discussion} we summarize our work, as well as discuss avenues for future work to build on these results.

For this work, we have assumed a $\Lambda$CDM cosmology with $\Omega_{\rm \Lambda} = 0.73$, $\Omega_{\rm M} = 0.27$, $\Omega_{\rm B} = 0.044$, $\sigma_8  = 1.7$, and $h = 0.71$.

\section{Methodology}\label{sec:Methods}

We use the moving-mesh code {\tt AREPO} \citep{springel10} for our simulations. We present two simulations with a $2.5$ Mpc box size, $768^3$ DM particles with mass M$_{\rm DM} = 1.1 \times 10^3 $ M$_\odot$, and $768^3$ Voronoi mesh cells with gas mass M$_{\rm B} = 200 $M$_\odot$, evolved from ${\rm z}=200$ to ${\rm z}=12$. Gas mesh cells in the simulation become eligible for collapse to form stars when their mass exceeds the Jeans mass on the scale of the cell. Subsequently, using the stochastic procedure described in \citet{Marinacci+19}, eligible gas cells are converted into star particles on the free-fall timescale. Star particles are implemented as collisionless particles with the mass of the gas cell that gave rise to them.

We use a modified version of {\tt CMBFAST} \citep{seljak96} to generate transfer functions for our initial conditions, incorporating first-order scale-dependent temperature fluctuations \citep{NB}, and the effect of the streaming velocity. Following the methods of \citet{chiou+19,Chiou+21, Lake+21, Nakazoto+22, Lake+22}, we generate initial conditions setting $\sigma_8 = 1.7$. This choice allows us to simulate a rare, overdense region where structure forms early, which increases our statistical power. This environment is similar to those that form galaxy clusters. As discussed in \citet{park+20}, this choice is also similar in effect to increasing the redshift of structure formation compared to the Universe overall by a factor of $\sqrt{2}$. We would, for example, expect corresponding structures to those in our simulation at ${\rm z}=20$ to form in a region with the bulk properties of the Universe at ${\rm z}\sim14$.

We present $2$ simulations in this paper, labelled as $2$v and $0$v. We use 2v and 0v here to indicate the stream velocity in the simulation. $2$v simulations use a streaming velocity of $2\sigma$ = 11.8 km s$^{-1}$ at the initial redshift ${\rm z}=200$, applied as a uniform boost to the x velocity of the baryons, as in \citet{popa}. $0$v runs use the same initial conditions, but do not include a streaming velocity.

\begin{figure*}[t]

\centering

\gridline{\fig{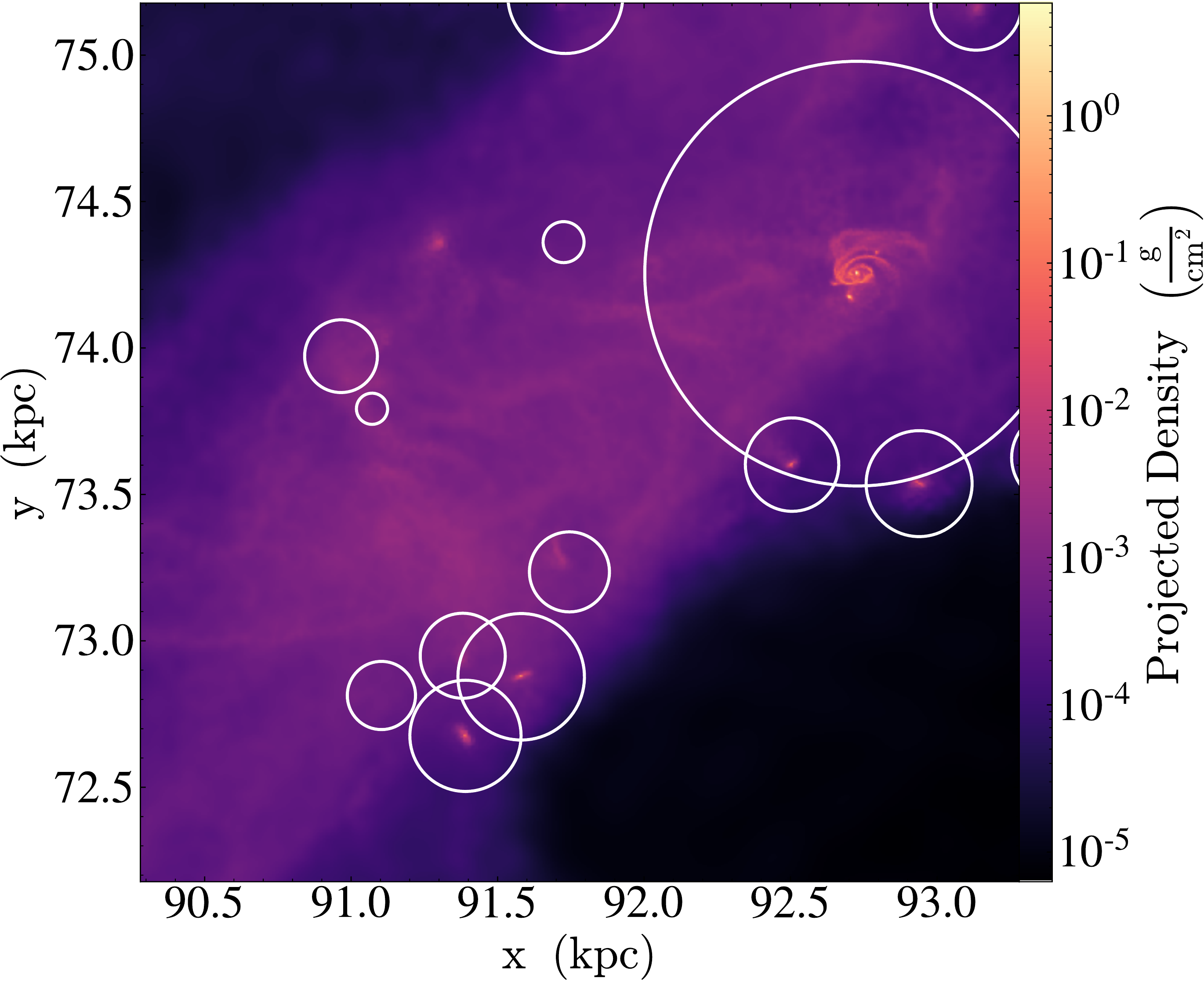}{0.33\textwidth}{\centering Gas Density without Streaming Velocity}
          \fig{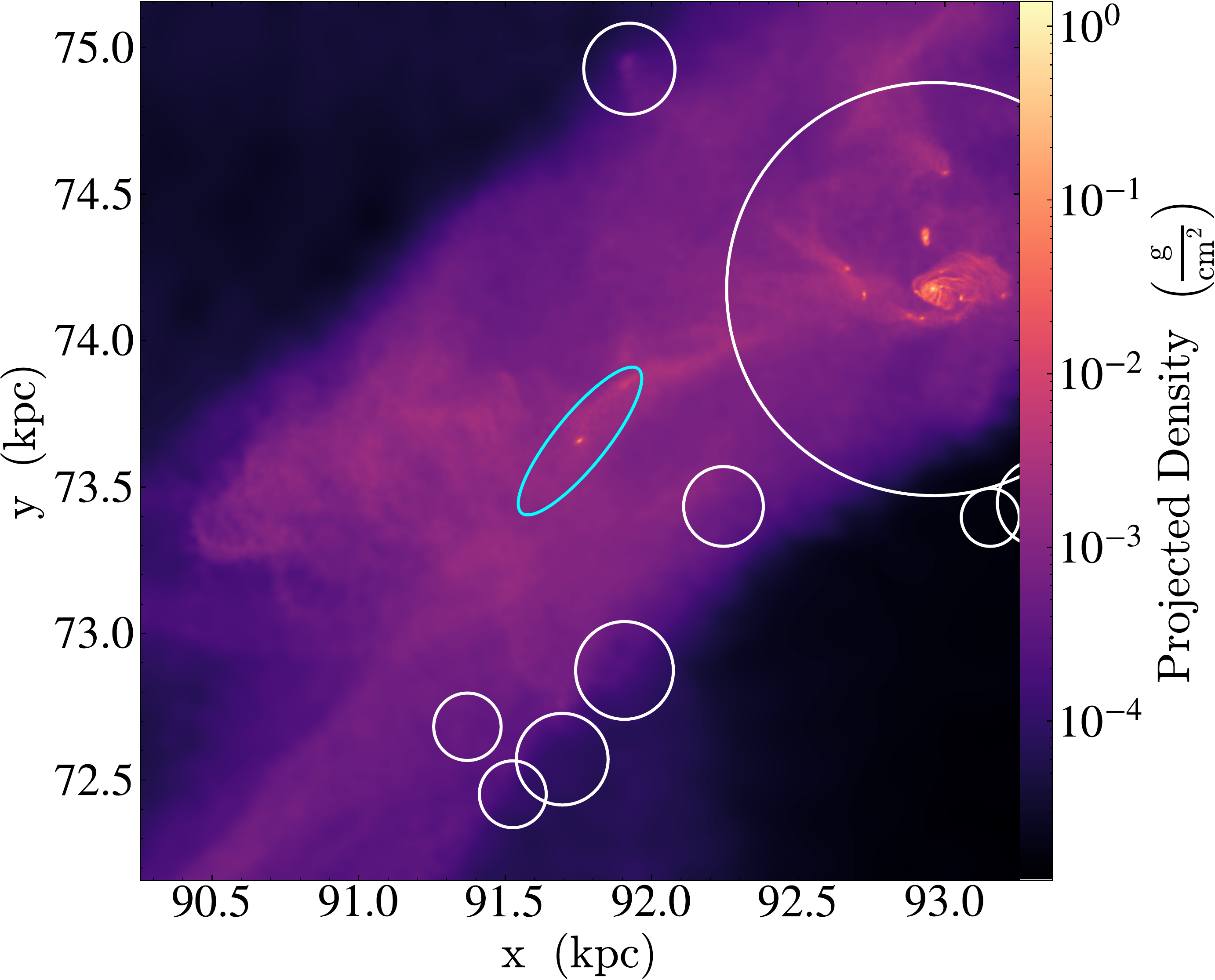}{0.33\textwidth}{\centering Gas Density with Streaming Velocity}
          \fig{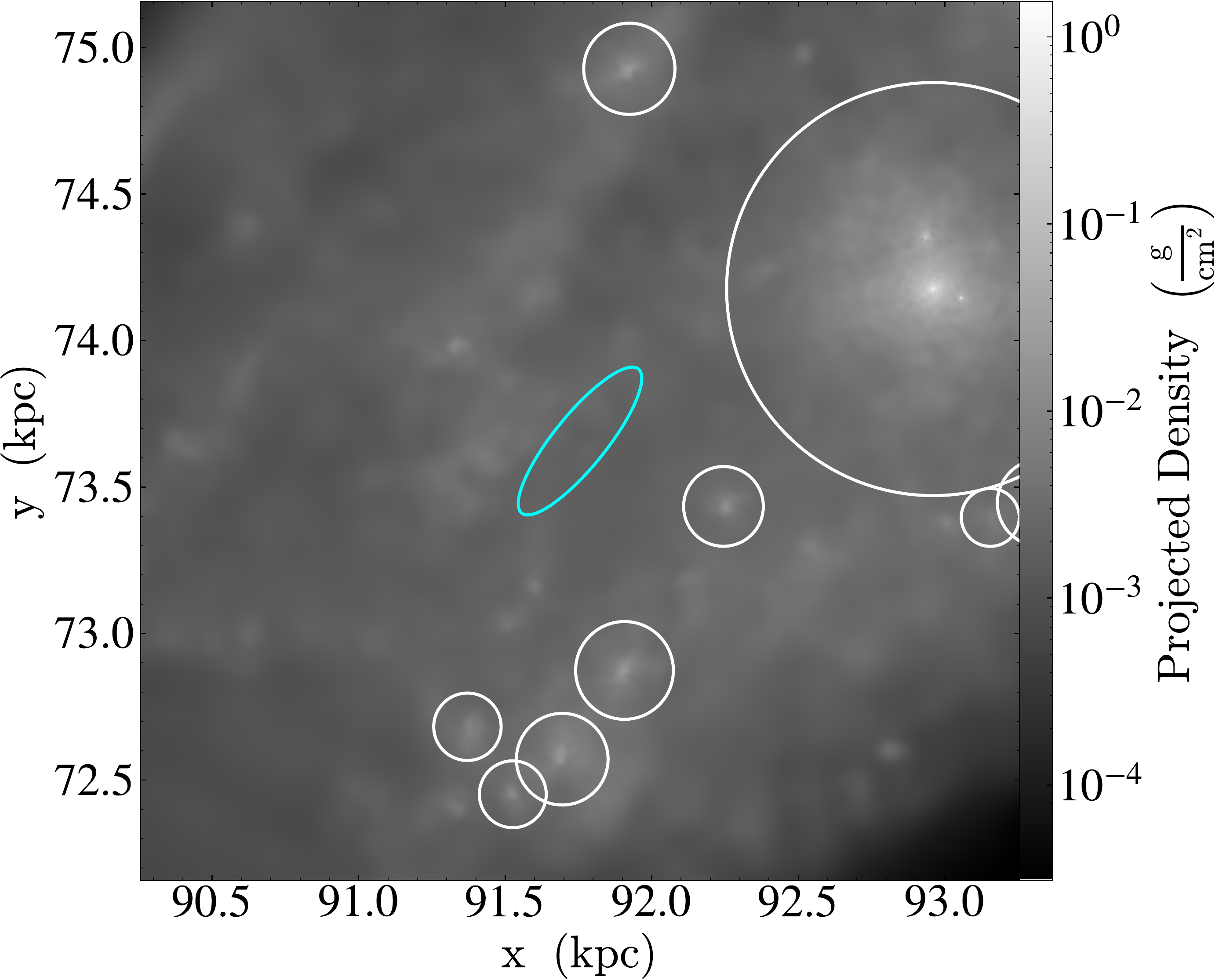}{0.33\textwidth}{\centering DM Density with Streaming Velocity}
          }
    \caption{The gas density field around a star-forming SIGO (marked with a cyan ellipse) and local dark matter halos (marked with white circles) in our simulations at ${\rm z}=15$. The color scale shows the column mass density of matter in the box to a depth of 3 kpc, centered on the center of mass of the SIGO in run $2$v. From left to right, we consider: \textbf{(a)} gas density without the streaming velocity ($0$v), \textbf{(b)} gas density with the streaming velocity ($2$v), and \textbf{(c)} the dark matter density with the streaming velocity ($2$v). All three panels show the same region. As can be seen in the middle panel, a SIGO is embedded in a larger shock (the high-density region of gas) and has a central high-density region/star formation site. The central SIGO here has a first generation of about $2\times10^4$ M$_\odot$ of stars by ${\rm z}=15$ in run $2$v. The SIGO does not exist in run $0$v (left).}\label{fig:ExampleSIGORho}%
\end{figure*}

We use the chemistry and cooling library {\tt GRACKLE} \citep{Smith+17, Chiaki+19} to track non-equilibrium chemical reactions and their associated radiative cooling explicitly in the gas. This includes molecular hydrogen and HD cooling, as well as chemistry for 15 primordial species: e$^-$, H, H$^+$, He, He$^+$, He$^{++}$, H$^-$, H$_2$, H$_2^+$, D, D$^+$, HD, HeH$^+$, D$^-$, and HD$^+$. The cooling rate of molecular hydrogen includes both rotational and vibrational transitions \citep{Chiaki+19}.

In this \textit{letter}, we use the object classifications from \citet{chiou18} to identify SIGOs and DM halos. DM halos are identified using an FOF algorithm with a linking length that is 20\% of the mean DM particle separation, or about 650 cpc. This algorithm calculates the location and virial radius of DM halos in the simulation output, assuming sphericity for simplicity \citep[although DM halos at these times are distinctly aspherical e.g.,][]{Sheth+01, Lithwick+11, Vogelsberger+11, Schneider+12, Vogelsberger+20}. The same FOF algorithm run on the gas component of the output then allows us to identify gas-primary objects. Star particles are associated with the gas-primary object that their nearest-neighbor gas cell belongs to. We require these objects to contain a minimum of 100 gas cells and star particles to be considered as SIGOs \citep{Chiou+21}. 

SIGOs are generally quite elongated in gas streams, so each gas-primary object is next fit to an ellipsoid, by identifying an ellipsoidal surface that encloses every particle in the object \citep{popa}. These ellipsoids are tightened by shrinking their axes by 5\% until either the ratio of the axes lengths of the tightened ellipsoid to those of the original ellipsoid is greater than the ratio of the number of gas cells contained in each, or 20\% of their particles have been removed, following \citet{popa}. We then apply a final filter requiring that SIGOs be located outside the virial radius of all DM objects, and have a gas+stars mass fraction of above 60\%, as in \citet{Nakazoto+22} and \citet{Lake+22} \footnote{Below the gas+stars mass fraction threshold of $60$\%, there are a variety of objects falsely identified by the FOF as SIGOs outside of halos in the no-streaming-velocity case (which may be nuclear star clusters, or compact gas objects in extended halos). At and above this threshold, many objects present in the run with the streaming velocity but absent in the run without it (true SIGOs) are removed, but very few objects falsely identified as SIGOs remain to be removed in the no-streaming-velocity run, as shown by \citet{Lake+22}.}. This limits false detections of SIGOs, as the filamentary nature of gas in runs with molecular hydrogen cooling enabled tends to result in the misidentification of SIGOs with a lower gas+stars fraction.

\section{Star Cluster Formation in SIGOs}\label{sec:Results}

In the classical description of structure formation (i.e., no streaming velocity) stars often form inside, and at the centers of, DM halos \citep[e.g.,][]{Tegmark+97}. However, the streaming velocity acts to separate gas and eventually stars from the centers of these halos \citep[e.g.,][]{TH,Williams+22}. In \citet{Lake+22}, we suggested that it may be straightforward for SIGOs to form stars. 
In this section, we analyze the evolution of a characteristic SIGO as it becomes a star cluster in the early Universe (z$\sim 17-12$).

\begin{figure*}[t]

\centering
   
\includegraphics[width=\textwidth]{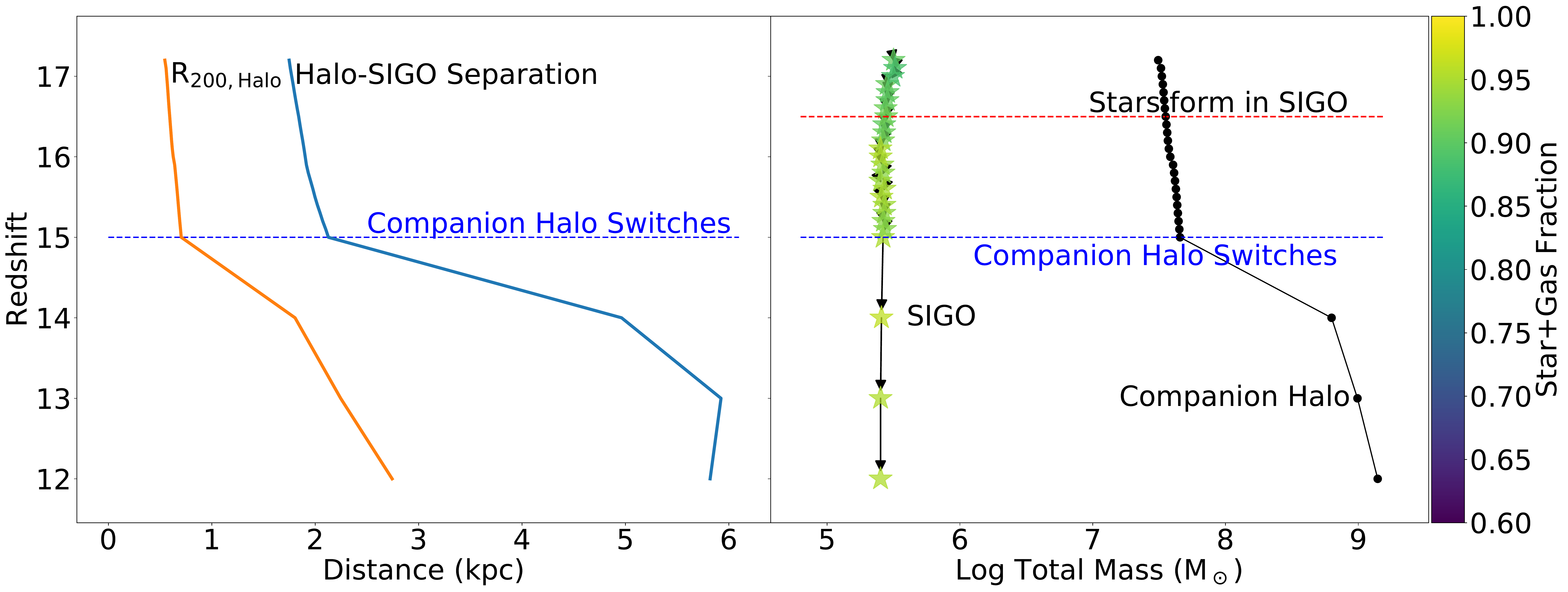} 
    
    \caption{{\bf The Evolution of a Star-Forming SIGO:} Here in the left panel we show the evolution of the $R_{200}$ of the companion halo to the SIGO in Figure \ref{fig:ExampleSIGORho} (defined as the nearest halo to the SIGO as a function of the halo's $R_{200}$). We also show the separation between the centers of mass of the SIGO and companion halo. In the right panel, we show the evolution of the total mass of the SIGO and the companion halo. The SIGO is indicated with a star symbol in this panel. It begins forming stars at $z\sim16.5$, as shown in the right panel. The black dots and lines in this panel indicate the mass evolution of the halo most closely associated with this SIGO at each redshift. The colors show the mass fraction of stars and gas within the SIGO compared to its total mass, showing that it maintains a high gas fraction throughout its evolution. Just after redshift $15$, the companion halo that had previously been closest to the SIGO, as a function of its $R_{200}$, is supplanted by a slightly more distant, but significantly larger, halo. This transition is marked with a blue horizontal line labelled "Companion Halo Switches" in both panels.
    }\label{Fig:SIGOEvolution}
\end{figure*}

\begin{figure*}[t]
\centering

\gridline{\fig{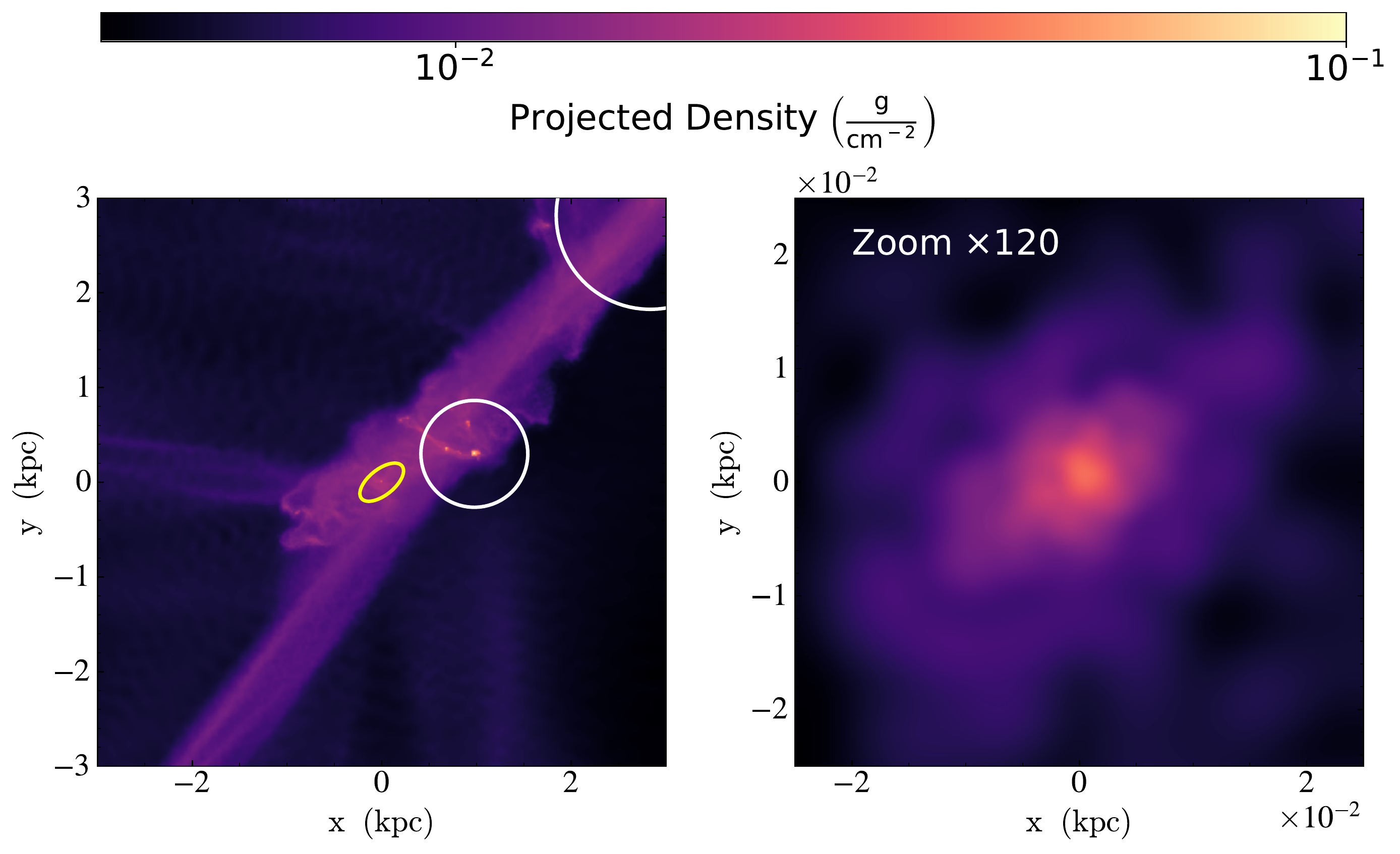}{.66 \textwidth}{${\rm z}=17$}
          }
\gridline{\fig{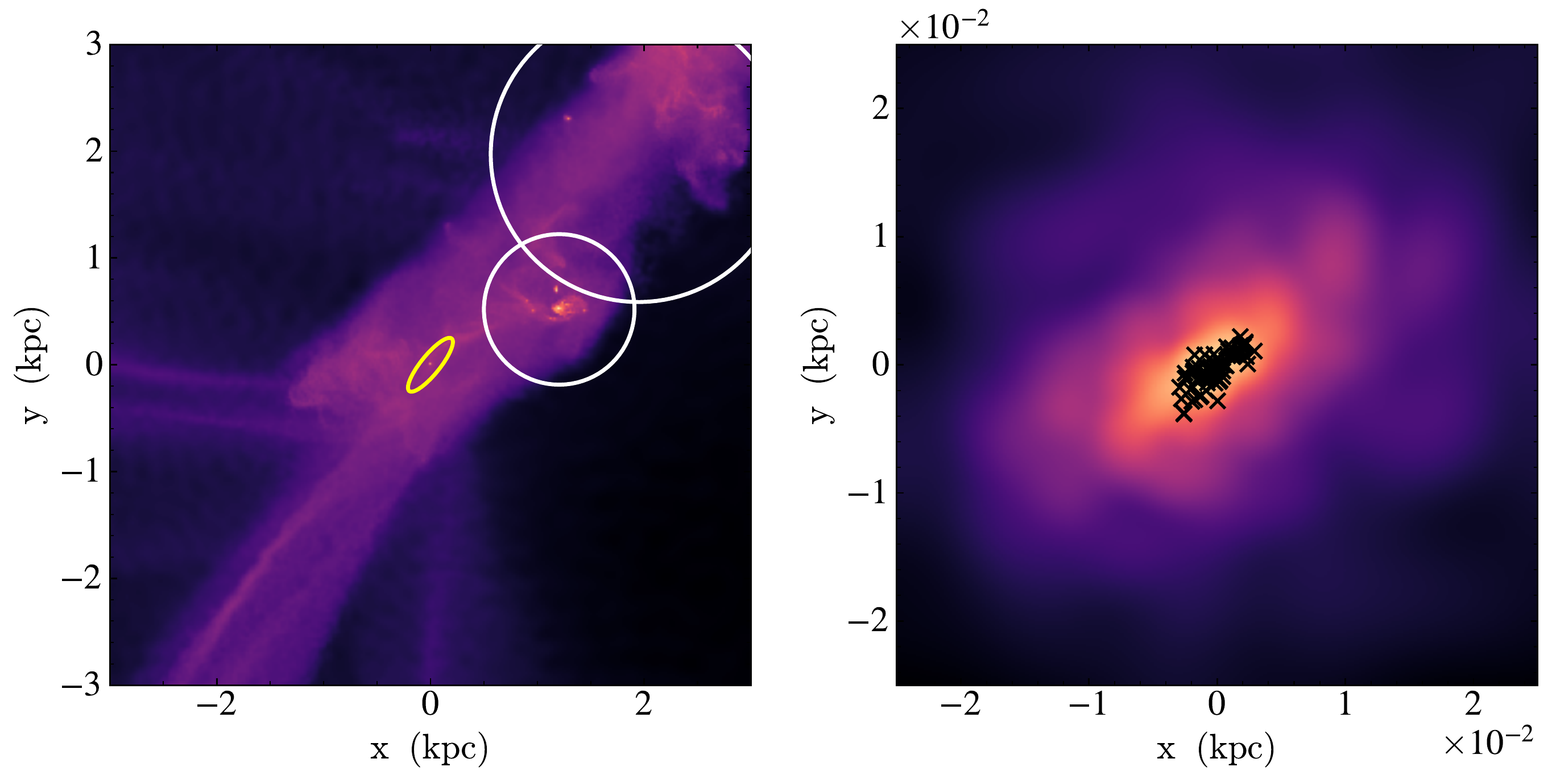}{.66 \textwidth}{${\rm z}=15$}
          }
\gridline{\fig{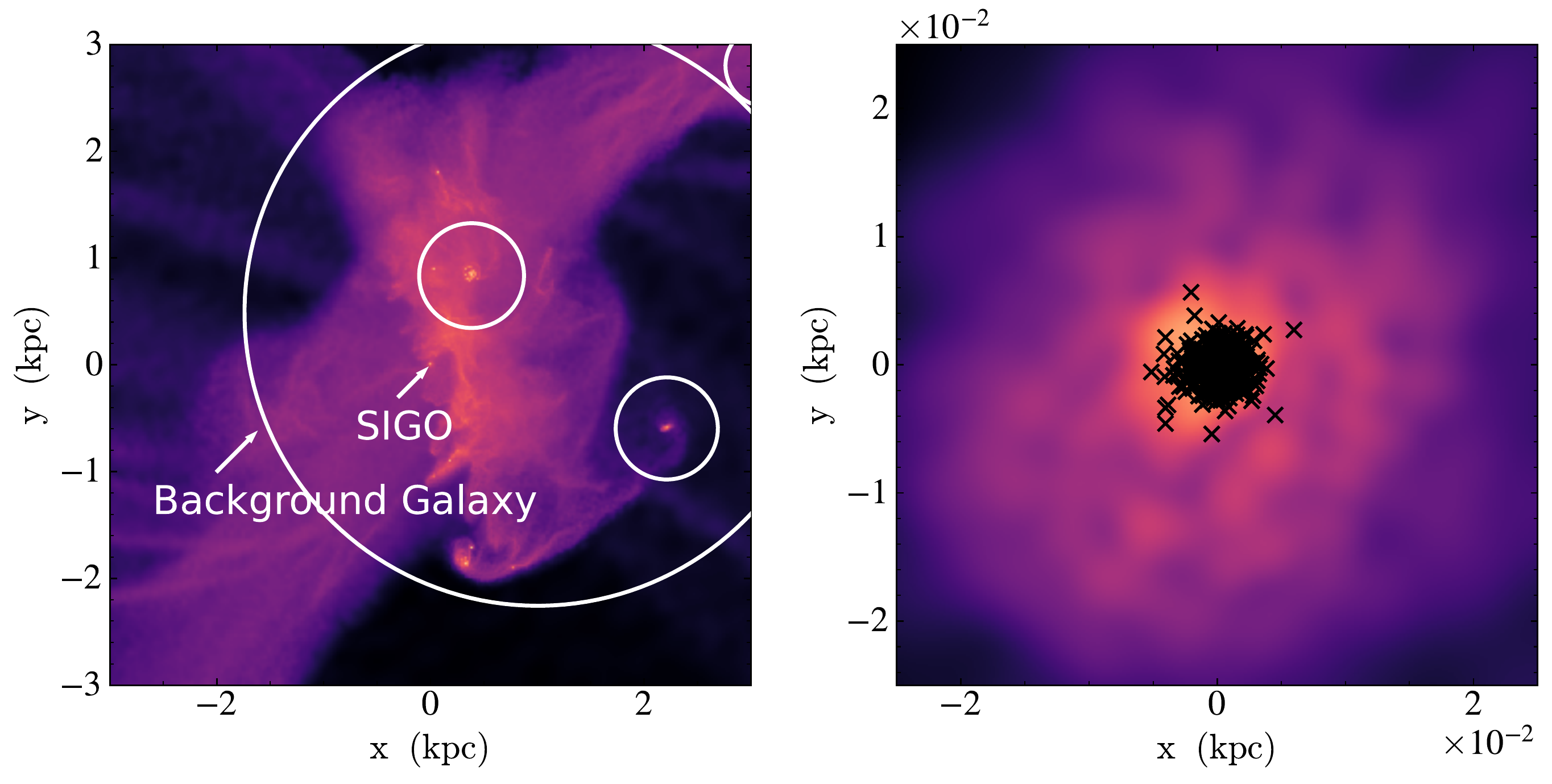}{.66 \textwidth}{${\rm z}=12$}
          }

\caption{\textbf{The Formation of a SIGO-Derived Globular Cluster: } Here we show successive snapshots of a SIGO (yellow ellipse) becoming a star cluster interacting with DM halos (white circles) in our $2\sigma$ streaming velocity run. Stars are marked in the right, zoom-in panel of each snap with black Xs. In the top panel at ${\rm z}=17$, the SIGO is outside any DM halo and has not formed stars. By ${\rm z}=15$ in the middle panel, the SIGO is undergoing star formation. In the bottom panel at ${\rm z}=12$, the SIGO (indicated by a white arrow) is no longer identified as an independent gas object, caught on the edges of a protogalaxy merger. This SIGO's fate is likely as a bound star cluster in this newborn galaxy (marked as background galaxy, and also referred to as a protogalaxy in the text). Two other smaller halos are also present in the bottom panel.}\label{fig:EvolutionCartoon}
\end{figure*}

Figure \ref{fig:ExampleSIGORho} shows this SIGO at the redshift (${\rm z}=15$) of its peak star formation in the two runs, i.e., the streaming velocity run ($2$v, middle and right panels) and the no streaming velocity run ($0$v, left panel) at ${\rm z}=15$. As shown, the SIGO is not present in the run without streaming velocity ($0$v, left panel). While a small gas overdensity is present near the top left of the frame, it is associated with a DM overdensity missed by the FOF and the baryon fraction within 0.15 kpc of the center of the gas overdensity is $22\%$, confirming that it is not a spurious object related to the SIGO. At all snapshots before and during star formation in the streaming velocity runs, the SIGO  contains less than $200\times$ the mean DM density. The SIGO begins forming stars at ${\rm z}\sim16$ in run $2$v. 
 
  As depicted in Figure \ref{fig:ExampleSIGORho}, the predominant DM halos exist in both runs. However, as mentioned, the SIGO only exists in the presence of the streaming velocity and is embedded in the gas stream \citep[e.g.,][]{Nakazoto+22,Lake+22,Williams+22}. The SIGO in Figure \ref{fig:ExampleSIGORho} contains $2\times10^4$~M$_\odot$ of stars in run $2$v at ${\rm z}=15$. It is also apparent that the gas is shifted between the left and middle panels, similarly as noted by \citet{Nakazoto+22} and \citet{Williams+22}.

The SIGO's mass evolution can be seen in the right panel of Figure~\ref{Fig:SIGOEvolution}, and its surface density evolution is shown in Figure~\ref{fig:EvolutionCartoon} in context. In Figure~\ref{Fig:SIGOEvolution}, the example SIGO (yellow-green star in the right panel) is associated with progressively larger nearby halos (in black), as the halos undergo a process of hierarchical mergers and accretion. The companion halo, defined as the nearest halo to the SIGO as a function of the halo's $R_{200}$, changes just after $z=15$, marked in the Figure with a blue horizontal line. The SIGO's mass is comprised of both gas and stars and is only slightly shrinking, due to the loss of some gas \citep[likely due to $2$-body interactions between particles at our simulation's limited resolution, e.g. ][]{Lake+22}. We note that the stars and gas are gravitationally bound. 

The left panel of Figure~\ref{Fig:SIGOEvolution} shows the evolution of the physical separation between the centers of mass of the SIGO and its companion halos (in blue), as well as the evolution of the $R_{200}$ of the SIGO's companion halos, defined as the radius around the halo that encloses $200$ times the critical density of the Universe. As one can see, the SIGO is slightly drifting away from its first companion halo at and before $z=15$, but the SIGO begins to fall into its second, much larger (and growing) companion halo by redshift $12$. At $z=12$, the SIGO is outside this halo at a center-of-mass separation $\Delta {\rm R} = 2.1 {\rm R}_{\rm 200, Halo}$, or about $6$~kpc.

 The SIGO and companion DM halos' evolutionary processes are visualized in the left column of Figure~\ref{fig:EvolutionCartoon}, which shows $6\times6$~physical kpc boxes, left column, and a zoomed-in region ($50\times50$~pc) in the right column.  The different rows show three redshifts: ${\rm z}=17, 15$, and ${\rm z}=12$.  As shown in this Figure, the nearby large DM halo at ${\rm z}=17$ is in the process of merging with a larger DM halo located at the right top corner of the image.  This tidally separates the SIGO from the nearby DM halo: we see this DM halo slightly further away from the SIGO at ${\rm z}=15$. This process may give the SIGO more time to cool and form stars outside of the halo before accretion, as well as limiting tidal forces from the halo on this SIGO. Stars start forming at ${\rm z}=16.5$ (see Figure~\ref{Fig:SIGOEvolution}), when the central $10$ pc has a gas surface density of about $880$ M$_\odot$ pc$^{-2}$. Significantly, tidal forces on this SIGO from its companion halo at this time are more than an order of magnitude smaller than the forces of self-gravity within this SIGO, allowing this SIGO's collapse to occur mostly unaffected by tides \citep[e.g.][]{Jog+13}. These results are subject to the exact configuration of a SIGO, and it may be possible that tides impact the collapse of other SIGOs, though that is beyond the scope of this paper.


Subsequently, the halo is accreted onto another nearby protogalaxy (also referred to as the second companion halo), soon after z=$12$. In the bottom panel of Figure \ref{fig:EvolutionCartoon}, the star cluster is located just outside of this large protogalaxy but has remained intact with the longest axis radius (from an ellipsoid fit) on the order of $6.6$~pc, and shortest axis of $3.4$~pc. This nascent star cluster is comprised in its entirety of stars that originated within the SIGO, and is now gravitationally bound to the $10^9$ M$_\odot$ (total mass) protogalaxy, which also has stars of its own. The star cluster contains no dark matter and has a stellar mass of $7.4\times10^4$ M$_\odot$. We subsequently refer to this as a globular-cluster-like object. 

In order to answer whether the SIGO is expected to be hosted by the protogalaxy, we ran a two-body simulation of the subsequent evolution of the SIGO and protogalaxy including cosmological expansion, confirming that the SIGO's orbital path enters the virial radius of the protogalaxy on a bound orbit. This analysis suggests that the SIGO is likely to fall within the virial radius of this protogalaxy within $100$~Myr of the end of our {\tt AREPO} hydrodynamic N-body simulation.

The end state of this SIGO as a cluster residing within a halo is commonplace. As shown in \citet{Lake+22}, star-forming SIGOs are expected to be accreted onto nearby halos shortly after forming stars. The hierarchical merging of these halos allows the most massive halos to accrete SIGO-derived star clusters. The protogalaxy which accretes this particular SIGO-derived star cluster is one of the largest protogalaxies in the simulation at all snapshots and could potentially host several SIGO-derived objects at later redshifts as it accretes nearby systems. On Gyr timescales, based on these results and those of \citet{Lake+22}, we expect more massive halos to host more SIGO-derived star clusters, gained through hierarchical formation. This particular cluster, as well as other low-mass star clusters derived from this process, eventually likely disperses through relaxation within its host halo on these Gyr timescales, while more massive clusters may survive \citep[e.g.,][]{naoznarayan14}.

An additional important property of this cluster is the degree of rotational support compared to random motion within its constituent stars. It is general consensus that local globular clusters are supported by random motion rather than ordered rotation, and this SIGO is not exceptional in that regard. Similarly to \citet{chiou18}, we express the spin parameter as 
\begin{equation}
    \lambda_{\rm SIGO} = \frac{J_*}{\sqrt{2}M_*v_{c}R_{\rm max}},
\end{equation}
where $M_*$ is the total stellar mass in the star cluster, $v_{c}$ is the circular velocity at a distance $R_{\rm max}$ from the center of the cluster, and $R_{\rm max}$ is the maximum axis radius of the star cluster determined by an ellipsoid fit. At $z=12$, this SIGO's star cluster has $\lambda_{\rm SIGO} \approx 0.072$, which is comparable to the spin parameter of many present-day globular clusters \citep[e.g.,][]{Kamann+18}.

Note that the streaming velocity for the $2$v case is $0.7$~km~s$^{-1}$ at z=$12$ (comparable to $59.5$~km~s$^{-1}$ at the time of recombination, and $1$~km~s$^{-1}$ at z$\sim17$, when the SIGO's overdensity initially formed). This streaming velocity injects turbulence into the gas, forming the SIGO \citep{Lake+22}. As such, one expects a signature of this turbulence to be left in the SIGO's velocity dispersion. The velocity dispersion of the stars in the cluster is estimated as $2.3$ km s$^{-1}$ at z=$12$  \citep[comparable to that of present-day globular clusters, e.g.,][]{Kamann+18}, suggesting that other sources of velocity dispersion also play major roles in the cluster.

\section{Detectability by JWST}

\begin{figure}[t]

\centering
\includegraphics[width=\linewidth]{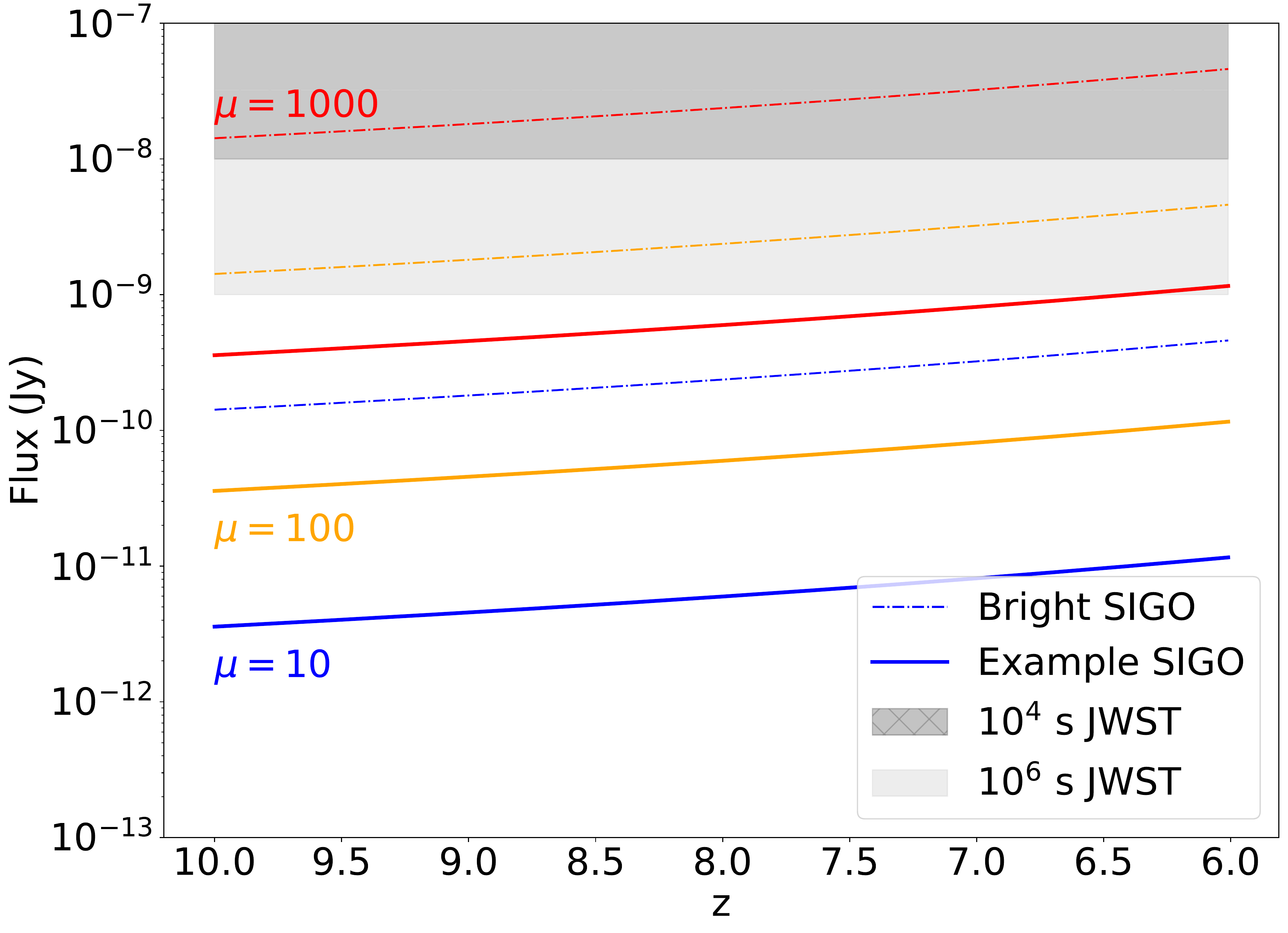} 
    
    \caption{{\bf SIGO flux as a function of redshift of initial star formation, and JWST detectability.} We consider the possibility of observing a lensed SIGO beginning star formation at later redshifts, similarly to the observation of a faint, lensed source in \citet{Welch+22}. Solid lines show the modeled 1500 \AA \space  flux  adopting the star formation rate estimate directly from our simulation (i.e., Figure~\ref{fig:SFRPlot}, in Appendix~\ref{sec:appendix}). However, as discussed in \citet{Lake+22}, larger SIGOs more readily form stars; therefore, we also consider a more massive, larger SIGO, with a gas mass of $\sim 10^7$~M$_\odot$. We scale the star formation rate linearly as a function of mass \citep[as implied from the SFR-mass relation, e.g.,][]{Lada+12}. This example is shown in dot-dashed lines, labelled as `Bright SIGO'. For the two examples we show three possible magnifications, of $10$ (blue),$100$ (orange), and $1000$ (red), from bottom to top.  The light (dark) shadowed regime shows characteristic JWST sensitivity after $10^4$ ($10^6$) seconds exposure time, with an SNR of about $10$, in agreement with the JWST Exposure Time Calculator \citep{Pontoppidan+16}.     
 }\label{Fig:Detectability}
\end{figure}

At high redshifts, one primary mechanism by which objects like these SIGO-derived star clusters could be detected is UV emission from young massive stars \citep[e.g.,][]{Sun+16,Hegde+23,Senchyna+23}. As a proof of concept examining whether SIGOs could be observed with JWST, we use a semi-analytical model based on the example SIGO in this simulation to examine the flux from these young, massive stars in SIGOs forming at later redshifts. As a proof of concept, we model the emission from these test SIGOs, assuming that they share the star formation rate of our example SIGO\footnote{Though note that the trend of star formation rates with redshift in SIGOs is not established, and could be impacted by decreasing gas densities with redshift.}, but form at various redshifts immediately prior to and during Reionization (${\rm z} = 6-10$). The UV luminosity of these SIGOs is given approximately by \citet{Sun+16}:
\begin{equation}\label{Eq:luminosity}
\dot{M}_{\rm SFR} = \mathcal{K}_{\rm UV,1500}\times L_{\rm UV,1500} \ ,
\end{equation}
where we take $\dot{M}_{\rm SFR}$ to be the average star formation rate in our example SIGO in the $20$ Myr period following the start of its star formation (after which, in a simulation with feedback, star formation may be quenched). See Appendix~\ref{sec:appendix} for further discussion of the effect of this time duration. $L_{\rm UV,1500}$ is the rest-frame UV luminosity at $1500$ \AA. $\mathcal{K}_{\rm UV,1500}$ here is a fiducial constant, which following \citet{Sun+16} we set to be approximately $\mathcal{K}_{\rm UV,1500} = 1.15\times10^{-28}$ ${\rm M}_\odot {\rm yr}^{-1} / {\rm ergs}$ ${\rm s}^{-1} {\rm Hz}^{-1}$, which assumes a Saltpeter IMF. It is important to note here that this constant most likely underestimates the luminosity at a given star formation rate, as SIGOs have extremely low metallicities and would be likely to have a top-heavy IMF.

The solid lines in Figure~\ref{Fig:Detectability} show the flux from test SIGOs with this star formation history, placed at varying redshifts. Three solid lines are shown, representing three different gravitational lens magnifications ($\mu=10,100$ and $1000$, bottom to top). Characteristic detectability from JWST at two exposure times is shown as gray-shaded regions. As seen in the Figure, an object with properties of our example SIGO, forming stars at the end of Reionization and magnified by a factor of a thousand or more, would only just be observable in a JWST field. UV emission from SIGOs similar to our example SIGO would likely not be observable with JWST at all. 

However, the SIGO we study in this \textit{letter} is likely not the most luminous possible SIGO, so to understand whether any SIGO would be detectable by JWST requires a model for the most luminous possible SIGO. The most massive SIGOs found have gas masses approaching $10^7$~M$_\odot$ \citep[][]{Lake+22}, exceeding the gas mass of this SIGO at the start of star formation by a factor of $40$. We argue that an approximate value for the star formation rate of the most luminous SIGO can, then, be given by multiplying the star formation rate of our example SIGO by this factor of $40$ ratio in the SIGOs' gas masses \citep[based on the SFR-mass relation, e.g.,][]{Lada+12}. This assumes that the star formation efficiency of SIGOs of different masses is the same.

We overplot the simulated observed flux from this characteristic most luminous SIGO in Figure~\ref{Fig:Detectability} with dashed lines. As one can see, in a very deep JWST field, strong lensing with a magnification of around $100$ or better may allow a particularly luminous SIGO to be observable even at very high redshift, given sufficient exposure time. This result is not surprising, in the context of the recent observations of even individual stars or star systems at high redshift in such lensed fields \citep[e.g.,][]{Welch+22}.

\section{Discussion}\label{sec:Discussion}

SIGOs (Supersonically Induced Gas Objects) are a natural consequence of early structure formation in $\Lambda$CDM \citep{naoznarayan14}. These gas objects form in the early Universe (${\rm z}\gtrsim 10$), with little to no dark matter, in the patches of the Universe with non-negligible streaming velocity between the dark matter and the baryons. Interestingly, it was recently suggested that our own Local Group formed in a region with a large streaming velocity \citep{Uysal+22}, implying that the small-scale structure in our vicinity was greatly impacted by the streaming velocity.  Investigating the star formation of these objects is critical for predicting future local and high redshift observations.  While previous studies expected that these objects would eventually form stars \citep[e.g.,][]{chiou+19,Nakazoto+22,Lake+22}, until now, no study investigated the formation of stars in these objects. 

Here we present, for the first time, a study of the outcomes of star formation in SIGOs. We estimate the stellar mass of SIGOs, and follow the evolution of a SIGO after star formation. We present a $2\sigma$ streaming velocity run $2$v, and a control run without the streaming velocity ($0$v) for comparison. See Figure \ref{fig:ExampleSIGORho} for a comparison of these runs.

We find that some SIGOs form stars. As expected \citep[see][]{Lake+22}, many other SIGOs accrete onto nearby DM halos prior to forming stars, forming diffuse structures akin to DM GHOSts \citep{Williams+22}. In total, out of 5325 unique SIGOs found at integer redshifts, 9 SIGOs formed stars outside of a DM halo before ${\rm z}=12$ \footnote{The latter number is visually verified, ensuring that the SIGO is outside of concentrations of dark matter missed by the FOF and that the candidate SIGO did not originate/form stars in the nucleus of a halo at earlier times. We also verify that the star particles formed within the SIGO.}, and others form DM GHOSt analogs or may form stars later \citep{Williams+22}. We explore the population and overall star formation efficiency in Lake et al. in prep.

Here we focused on an example SIGO as a detailed case study of the formation and evolution of a GC-like star cluster. Figure \ref{fig:ExampleSIGORho} depicts this process in context, showing that simulation runs with the streaming velocity effect form a gas object in a location where there is none in the no-stream-velocity runs, and that the object is capable of cooling to form stars while fully outside of nearby DM halos.

Figure \ref{fig:EvolutionCartoon} depicts the evolution of a SIGO and its vicinity as it forms stars. By ${\rm z}=12$, the resultant GC-like star cluster is bound to a large, $10^9$~M$_\odot$ protogalaxy (although it lies beyond its ${\rm R}_{200}$, as shown by Figure~\ref{Fig:SIGOEvolution}). Based on a simple two-body simulation, the SIGO is likely to merge with the halo to form a bound cluster within the halo on a timescale of about $150$~Myr following the end of our {\tt AREPO} hydrodynamic N-body simulation.

The SIGO formed stars at ${\rm z}=16.5$ while still outside its closest DM halo. The SIGO was separated from this nearby halo through a tidal interaction with a third, larger halo, possibly allowing it to remain outside of the nearby halo as it formed stars. 
Subsequent evolution may yield a cluster that resembles a cluster resultant from a more classical evolution \citep[e.g.][]{Sameie+22}. The SIGO-derived clusters may have some differing characteristics, such as their velocity profiles \citep[e.g.,][]{Williams+22}. The resulting cluster could also have a high galactocentric distance compared to clusters that formed locally, owing to its accreted nature.

Note that in this study we do not include feedback or the effects of metal enrichment on the SIGO. Although radiative feedback will act to reduce the efficiency and duration of star formation within the SIGO, even low levels of metal enrichment can significantly increase cooling rates within it \citep[such as from pair-instability supernovae in nearby halos, ][]{Schauer+21}. In addition, the limited mass resolution of these simulations results in a significant underestimation of the effectiveness of molecular hydrogen cooling in SIGOs \citep{Nakazoto+22}. Furthermore, our softening scale of $2.2$~pc is a significant fraction of the radius of the star cluster, which is expected to lower our simulated star formation rates, acting together with our underestimation of the effectiveness of molecular hydrogen cooling to make star formation in SIGOs seem slower than it is. Taken together, these processes will have an ambiguous effect on the stellar mass of SIGOs. Higher stellar masses will increase both the longevity of star clusters derived from SIGOs as well as the potential for binary black hole mergers and gravitational wave events as these clusters evolve.

Already, early JWST observations have found potential candidates for high-z clusters, some with high galactocentric distances \citep[e.g.][]{Pascale+22,Senchyna+23}. Figure \ref{Fig:Detectability} explores this possibility, considering whether a SIGO-derived star cluster with properties similar to the cluster explored in this letter would be observable by JWST in the Reionization epoch and whether a SIGO that is especially massive and bright may be observable in the same epoch. We find that a SIGO with properties similar to the one in this letter  
would likely be unobservable with JWST (even considering lensing). However, a more massive SIGO similar to the most massive SIGOs in  simulations may be detectable via lensing, presenting the possibility of direct detections.

\acknowledgments

The authors thank the anonymous reviewer for their thoughtful comments that have substantially improved the paper. W.L., S.N., Y.S.C, B.B., F.M., and M.V. thank the support of NASA grant No. 80NSSC20K0500 and the XSEDE AST180056 allocation, as well as the Simons Foundation Center for Computational Astrophysics and the UCLA cluster \textit{Hoffman2} for computational resources. S.N thanks Howard and Astrid Preston for their generous support. B.B. also thanks the the Alfred P. Sloan Foundation and the Packard Foundation for support. MV acknowledges support through NASA ATP grants 16-ATP16-0167, 19-ATP19-0019, 19-ATP19-0020, 19-ATP19-0167, and NSF grants AST-1814053, AST-1814259,  AST-1909831 and AST-2007355. NY acknowledges financial support from JST AIP Acceleration Research JP20317829. YN also acknowledges funding from JSPS KAKENHI Grant Number 23KJ0728.

\appendix
\section{Effect of Time on Cumulative Star Formation Rate}\label{sec:appendix}

\begin{figure}[t]
\begin{centering}
\includegraphics[width=0.5\textwidth]{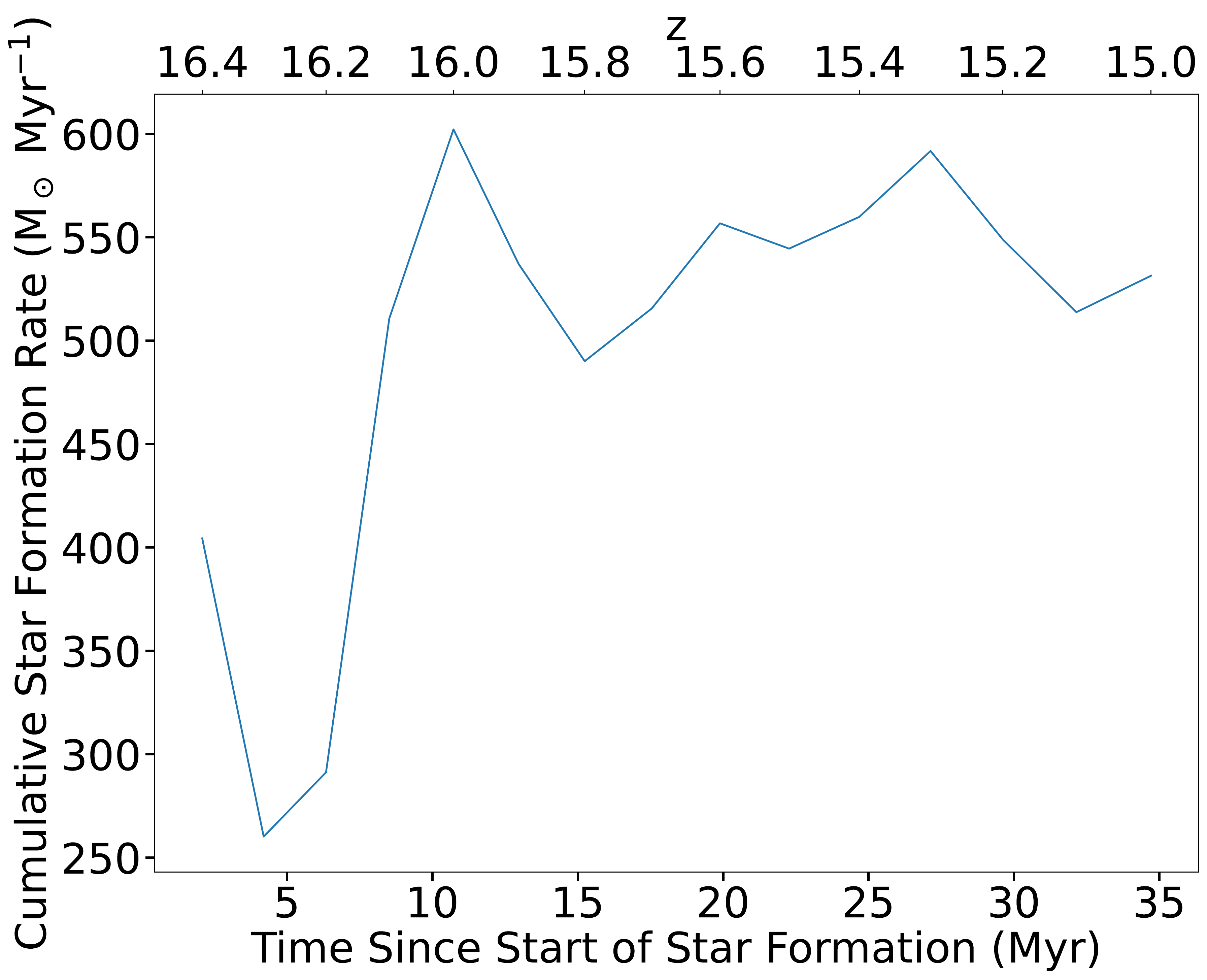}
\caption{Plot of the cumulative star formation rate (calculated as total stellar mass divided by time since the start of star formation) in the SIGO as a function of time. Regardless of time period considered, the star formation rate within the SIGO is the same to within a factor of about $2$.}\label{fig:SFRPlot}
\end{centering}
\end{figure}

We estimate the star formation rate directly from the rate at which star particles are formed in our simulation. This is shown in Figure \ref{fig:SFRPlot}. We note that we do not include radiative feedback, which eventually acts to quench star formation. As mentioned in Section~\ref{sec:Discussion}, the effects of feedback on our simulation may be balanced on timescales shorter than the quenching timescale by our cooling mechanism, which serves as a lower bound to the true cooling rate in SIGOs for the reasons given in the aforementioned section. The likely quenching timescale is $\sim10-20$~Myr, so in order to estimate the star formation rate in the SIGO during its initial starburst, we need to select a star formation rate over this period. As shown in Figure~\ref{fig:SFRPlot}, the star formation rate over this period is not a large source of uncertainty, as it varies only by about a factor of two over the timespan. As such, we use a period of $20$~Myr following the start of star formation in the SIGO to calculate a star formation rate, $557$ ${\rm M}_\odot$ ${\rm Myr}^{-1}$.


\bibliography{myBib}{}
\bibliographystyle{aasjournal}



\end{document}